\documentclass[12pt,preprint]{aastex}
\shorttitle{Bifurcation periods in LMXBs} \shortauthors{MA \& Li.}

\begin{document}


\title{The bifurcation periods in low-mass X-ray binaries: the effect of magnetic braking
and mass loss}

\author{Bo Ma and Xiang-Dong Li}
\affil{Department of Astronomy, Nanjing University, Nanjing 210093,
China} \email{xiaomabo@gmail.com, lixd@nju.edu.cn}


\begin{abstract}
The bifurcation period in low-mass X-ray binaries is the initial
orbital period which separates the formation of converging systems
(which evolve with decreasing orbital periods until the donor
becomes degenerate) from the diverging systems (which evolve with
increasing orbital periods until the donor star loses its envelope
and a wide detached binary is formed). We calculate systematically
the bifurcation periods of binary systems with a $1.4M_\sun$ neutron
star and a $0.5-2 M_\sun$ donor star, taking into account different
kinds of magnetic braking and mass loss mechanisms. Our results show
that the saturated magnetic braking can considerably decrease the
values of bifurcation period compared to the traditional magnetic
braking, while the influence of mass loss mechanisms on bifurcation
periods is quite weak. We also develop a semi-analytical method to
compute the bifurcation period, the result of which agrees well with
the numerical method in the leading order.
\end{abstract}


\section{Introduction}


One interesting and important topic in the secular evolution of
low-mass X-ray binaries (LMXBs) is the so-called ``bifurcation
period" $P_{\rm bif}$, the initial binary orbital period which
separates the formation of converging systems (which evolve with
decreasing orbital periods until the donor becomes degenerate) from
the diverging systems (which evolve with increasing orbital periods
until the donor star loses its envelope and a wide detached binary
is formed) \citep{tutukov85}. The first systematic investigations on
the bifurcation period were done by \citet{pylyser88,pylyser89}.
Neglecting mass loss from the binary system and assuming angular
momentum loss due to magnetic braking \citep[MB;][]{verbunt81} and
gravitational radiation \citep[GR;][]{landau}, these authors found
that the bifurcation period is in the range $P_{\rm bif}\sim
0.4-0.7$ day for LMXBs, and strongly depends on magnetic braking
efficiency. \citet{ergma98} included mass loss from the binary
system and re-calculated the bifurcation period for two mass
configurations ($M_1/M_\sun$, $M_2/M_\sun)= (1.4,1)$ and $(1.4,1.5)$
and two chemical compositions ($Z=0.003$, $0.03$). They pointed out
that the mass loss from the binary system also plays an impotent
role besides magnetic braking in determining the value of $P_{\rm
bif}$, while the chemical composition could only cause small change
in $P_{\rm bif}$. Their bifurcation periods are $P_{\rm bif}\sim
0.85-1.05$ day under conservative mass transfer, and $1.6-1.7$ times
larger if moderate non-conservative mass transfer is assumed.
\citet{pod02} found a bifurcation period around $18$ hr for a
$1.4M_\sun$ NS and a $1M_\sun$ companion star, where they defined
the bifurcation period as the orbital period when the Roche lobe
overflow just began, instead of the initial orbital period.

\citet{sluys05a,sluys05b} also investigated the bifurcation period
in LMXBs focusing the formation of ultra-compact X-ray binaries
(UCXBs), and specified the bifurcation period as ``the longest
initial period that leads to UCXBs within a Hubble time ($13.7$
Gyr)". UCXBs are bright X-ray sources with very short orbital
periods ($P \la 1$ h). The donor has to be a compact source like a
white dwarf or a compact core of an evolved giant star to fit in the
small Roche lobe size. Such sources may be formed through dynamical
processes including stellar collisions and common envelop evolution
\citep{clark75,rasio00,lombard06}. An alternative scenario for the
formation of such sources is through stable mass transfer in X-ray
binaries with a low- or intermediate-mass donor star, which may
explain the negative derivative of the 11-min source in NGC 6624
\citep{klis93,chou01}. It has been found that systems with initial
orbital period just below the bifurcation period may form UCXBs
\citep{nelson86,tutukov87,pylyser88,pod02,sluys05a}. \citet{pod02}
showed that the closer the initial orbital period to the bifurcation
period from below, the smaller the minimum orbital period will be
achieved. So the value of bifurcation period is crucial to
understanding the formation of UCXBs \citep{sluys05b}.

In this paper we make a systematic investigation on the bifurcation
period for binary systems containing an NS with a main-sequence (MS)
companion of mass from $0.5M_\sun$ to $2M_\sun$. This work was
motivated by recent progress in studies on mass and angular momentum
loss mechanisms in LMXB evolution. In previous works the MB law
originally postulated by \citet{verbunt81} and \citet{rappaport83}
was usually adopted. However, this law predicts too fast spin-down
of low-mass MS stars, contradicted with the observation of rapid
rotators in young open clusters \citep{sills00,andronov03}.
Obviously a modification of the MB law will have significant
influence on the period evolution  \citep{sluys05b}. Additionally, there is strong
evidence that during LMXB evolution the mass transfer is highly
non-conservative. Recent measurements of the masses of binary and
millisecond pulsars indicate that a large fraction of the
transferred mass may be lost from the systems rather accreted by the
NS \citep[][and references therein]{bassa06,steeghs07}\footnote{The
massive ($\sim 1.9 M_{\sun}$) NS discovered in the globular cluster
M5 \citep{freire08} may reflect a bimodal  distribution of the
initial masses of NSs (rather heavy accretion during the previous
LMXB evolution), as already predicted by hydrodynamical core
collapse simulations \citep{timmes96}.}. Theoretically, possible
ways of mass loss have been suggested, including ``evaporation" of
the donor \citep{ruderman89} or ``radio-ejection" of the transferred
material \citep{burderi01,burderi02,antona06} due to the pulsar
radiation/wind impinging on. In the latter case, the matter is lost
from the system at the inner Lagrangian ($L_1$) point, carrying away
angular momentum and altering the period evolution.

This paper is organized as follows. \S 2 briefly describes the
stellar evolution code, the binary models, and the physical
assumptions, especially the MB laws and the mass loss mechanisms.
Then we present the calculated results in \S 3. Our discussion and
conclusions are given in \S 4.


\section{Evolution code and binary model}
\subsection{The stellar evolution code}
We use an updated version of the stellar evolution code originally
developed by \citet[][see also Han et al. 2004, Pols et al.
1995]{eggleton71,eggleton72} to calculate the evolutions of binaries
consisting of an NS (of mass $M_1$) and an MS secondary (of mass
$M_2$). For the secondary star we assume a solar chemical
composition ($X=0.70$, $Y=0.28$, and $Z=0.02$), the ratio of mixing
length to pressure scale height $\alpha= 2.0$, and convective
overshooting parameter to be $0.12$. The opacity table is from
\citet{rogers92}, \citet{alexander94} and \citet{hubbard69}. The
effective radius of the Roche lobe for the secondary is taken from
\citet{eggleton83},
\begin{equation}
R_{\rm{L,2}}=\frac{0.49q^{-2/3}}{0.6q^{-2/3}+\ln(1+q^{-1/3})}a
\end{equation}
where $q=M_2/M_1$ is the mass ratio, and $a$ is the orbital
separation. Mass transfer rate via Roche lobe overflow is evaluated
as $-\dot{M}_2=RMT\cdot max(0,(R_2/R_{\rm
L,2}-1)^3)M_{\sun}$yr$^{-1}$, and we adopt $RMT=10^3$ in the
calculations.

\subsection{Mass and Angular momentum loss mechanisms}
For LMXBs the timescale of tidal synchronization is much shorter
than the characteristic evolutionary timescale of the binary, so we
can assume that the spin of the secondary star and the binary
orbital revolution are always synchronized. Assuming rigid body
rotation of the secondary star and neglecting the spin angular momentum
of the neutron star, the total angular momentum of the binary system
can be expressed as
\begin{eqnarray}
J&=&I_2\omega+J_{\rm orb} \nonumber \\
&=&I_2\omega+G^{2/3}M_1M_2(M_1+M_2)^{-1/3}\omega^{-1/3}
\end{eqnarray}
where $I_2$ is the moment of inertia of the secondary star, $\omega$
is the angular velocity of the binary.

We consider three kinds of mechanisms of angular momentum loss. The
first is the angular momentum loss due to gravitational radiation
\citep{landau}
\begin{equation}
\frac{{\rm d}J_{\rm{GR}}}{{\rm d}t}=-\frac{32}{5}\frac{G^{7/2}}{c^5}
\frac{M_1^2M_2^2(M_1+M_2)^{1/2}}{a^{7/2}},
\end{equation}
where c is the light speed. This mechanism is important only in
very short period binary systems.

The second angular momentum loss mechanism is for non-conservative
mass transfer. We assume that a fraction $\alpha$ of the transferred
mass is accreted by the NS, and the remaining mass is ejected out of
the binary as isotropic winds from the NS, carrying away the
specific angular momentum of the NS,
\begin{equation}
\frac{{\rm d}J_{\rm{ML}}}{{\rm
d}t}=-(1-\alpha)\dot{M}_2(\frac{q}{1+q})^2a^2\omega.
\end{equation}
In our numerical calculations we have set $\alpha=0$. Alternatively,
if the NS is spun up to be a millisecond pulsar, its radiation
pressure may be strong enough to halt the transferred matter at the
$L_1$ point and quench the accretion. This ``radio ejection'' may
cause almost all the matter from the secondary to be lost from the
binary \citep{burderi01,burderi02}. The corresponding rate of
angular momentum loss is
\begin{equation}
\frac{{\rm d}J_{{\rm ML}}}{{\rm d}t}=-\dot{M}_2 a_{L1}^2\omega
\end{equation}
where $a_{L1}$ is the distance from the $L_1$ point to the center
of mass of the binary system.

The third angular momentum loss mechanism is MB. For a low-mass
MS star with deep convection zone, stellar winds which are magnetically
coupled with the star can decelerate the stellar spin efficiently, thus
carrying away the orbital angular momentum because of tidal synchronization.
The widely used formula for such MB effect was postulated by
\citet{verbunt81} and \citet{rappaport83} as
\begin{equation} \label{oldmb}
\frac{{\rm d}J_{\rm{MB}}}{{\rm d}t}=-3.8\times10^{-30}M_2R_2^4\omega^3 \,\rm{dyn\,cm}.
\end{equation}
However, observations of rapid rotators in young open clusters
suggest a modification of the MB law at high rotation rate
\citep{sills00},
\begin{eqnarray}\label{newmb}
\frac{{\rm d}J_{\rm{MB}}}{{\rm d}t} & = & -K \omega^3
\left(\frac{R_2}{R_{\sun}}\right)^{0.5}
\left(\frac{M_2}{M_{\sun}}\right)^{-0.5},  \hspace{20pt} \omega \leq \omega_{\rm{crit}}, \nonumber \\
 \frac{{\rm d}J_{\rm{MB}}}{{\rm d}t} & = & -K \omega_{\rm{crit}}^2 \omega
\left(\frac{R_2}{R_{\sun}}\right)^{0.5}
\left(\frac{M_2}{M_{\sun}}\right)^{-0.5},  \ \omega >
\omega_{\rm{crit}},
\end{eqnarray}
where $K=2.7\times10^{47}$ gcm$^2$s \citep{andronov03},
$\omega_{\rm{crit}}$ is the critical angular velocity  at which the
angular momentum loss rate reaches a saturated state, given by
\citep{krishnamurthi97}
\begin{equation}
\omega_{\rm{crit}}(t) = \omega_{\rm{crit}\sun} \frac{\tau_{\rm{t}_0,\sun}}{\tau_{\rm{t}}},
\end{equation}
where $\tau_{\rm{t}_0,\sun}$ and  $\tau_{\rm{t}}$ are the global
turnover timescales for the convective envelope of the Sun at its
current age and of the secondary star at age $t$, respectively. They
can be calculated by integrating the inverse local convective
velocity over the surface convective envelope \citep{kim96}:
\begin{equation}
\tau_{\rm t}=\int_{R_b}^{R_2}\frac{{\rm d} r}{v},
\end{equation}
where $R_b$ is the radial distance from the center of the star to
the bottom of the surface convective envelope, and $v$ is the local
convective velocity from mixing-length theory \citep{bohm58}. Our
calculation gives $\tau_{\rm{t}_0,\sun}\simeq 28.4$ d, slightly
larger than $\tau_{\rm{t}_0,\sun}\simeq 13.8$ d  in
\citet{sluys05b}, but consistent with the results of \citet{kim96}
and \citet{jung07}.  See \citet[page 46]{eggleton06} for the discussion 
of a possible reason for different values of $\tau_{\rm{t}_0,\sun}$ calculated.

Following the suggestion of \citet{pod02}, we also add an ad hoc
factor
\[
\exp(-0.02/q_{\rm{conv}}+1) \; \rm{if}\;q_{\rm{conv}}<0.02,
\]
in Eqs.~(6) and (7), where $q_{\rm{conv}}$ is the mass fraction of
the surface convective envelop, to reduce the MB effect when the
convective envelope  becomes too small.

\subsection{Binary models}
To examine the influence of mass and angular momentum loss
mechanisms on the period evolution, we construct four models with
various mass and angular momentum loss combinations: (1) model 1 -
conservative mass transfer with traditional MB law (Eq.~[6]); (2)
model 2 - conservative mass transfer with saturated MB law
(Eq.~[7]); (3) model 3 - non-conservative mass transfer with mass
loss from $L_1$ point (Eq.~[5]) and saturated MB law (Eq.~[7]); and
(4) model 4 - non-conservative mass transfer with mass loss from the
NS (Eq.~[4]) and saturated MB law  (Eq.~[7]). In all the four
models, the initial NS mass is set to be $M_{1, {\rm i}}=1.4M_\sun$,
and the initial mass of the secondary  $M_{2,{\rm i}}$ ranges from
0.5 to $2.0M_\sun$.

\section{Numeric Results}
\subsection{The bifurcation periods}
Throughout this paper we define the bifurcation period $P_{\rm bif}$
as the initial binary orbital period $P_{\rm i}$ with a zero-age
main-sequence (ZAMS) companion star that separates converging from
diverging systems. We use $P_{\rm f}$ to denote the final orbital
period after the mass transfer. Another definition of the
bifurcation period used by \citet{pod02} is the orbital period when
the Roche lobe overflow just begins, which is expressed as $P_{\rm
rlof}$ in this paper.

\begin{figure}
\epsscale{.80} \plotone{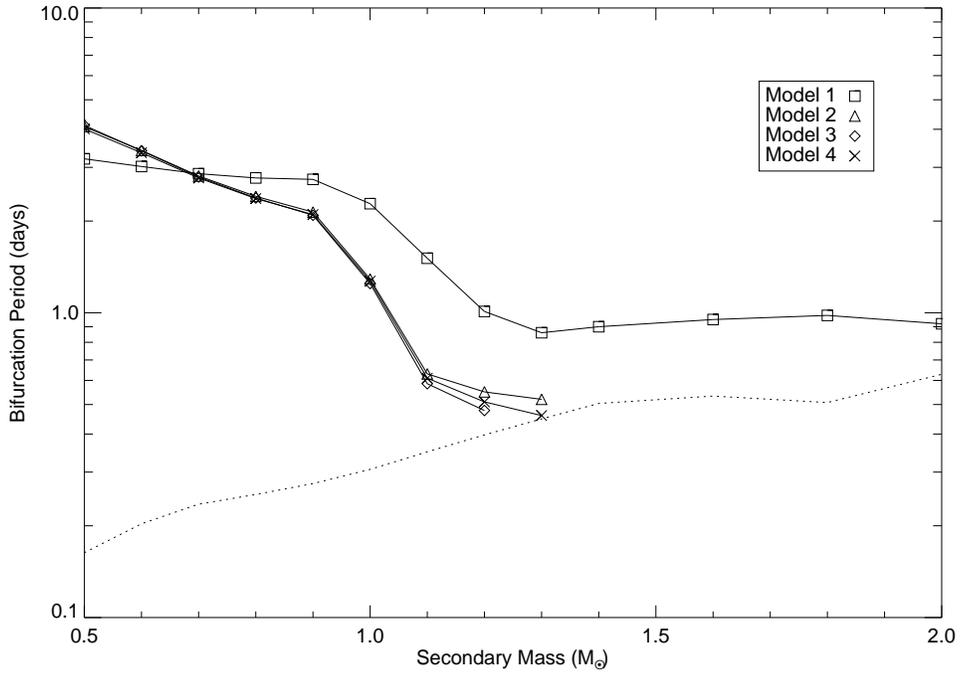} \caption{Bifurcation periods as a
function of the secondary mass in an LMXB for the four kinds of
models described in \S 2.3. The dotted line shows the minimum
initial period $P_{\rm ZAMS}$ that corresponds to a Roche lobe
filling zero-age main-sequence secondary star.\label{bifur}}
\end{figure}

The results of the bifurcation periods for the four models described
in \S 2.3 are summarized in Fig.~\ref{bifur} and Table~\ref{tab1}.
We also draw the minimum initial period $P_{\rm ZAMS}$ that
corresponds to a lobe-filling ZAMS donor star in Fig.~\ref{bifur}.
Several features are noted for the bifurcation periods in
Fig.~\ref{bifur}. First, the bifurcation periods for all the four
models decrease with increasing initial secondary mass from
$0.5M_\sun$ to $1.3M_\sun$. Second, in models with saturated MB,
there exists an upper limit of the initial secondary mass, beyond
which no bifurcation period exists. This upper limit is in the range
$\sim 1.2-1.3\; M_\sun$ for model 3, and $\sim 1.3-1.4\; M_\sun$ for
models 2 and 4. Third, comparing the bifurcation periods of model 1
with those of models $2-4$ indicates that the MB law plays the most
important role in determining the values of the bifurcation periods:
different MB laws can change the bifurcation periods by as much as
$\sim 60\%$, compared to $\sim 14\%$ (see Table~\ref{tab1}) caused
by different mass loss mechanisms.

In Table~\ref{tab1} we also present $P_{\rm rlof}$ following
\citet{pod02}. In model 1, we get $P_{\rm rlof}\simeq 18.3$ hr for
$M_{2,i}=1M_\sun$, which is in close line with the result of
\citet{pod02} ($17.7$ hr with $Z=0.001$, $Y=0.27$), where the difference 
could be explained as the difference between the metallicities we used. 
When we use saturated MB, $P_{\rm rlof}$ decreases to $\sim 11$ hr.

According to the calculated orbital period evolutions, LMXBs can be
classified into three categories: the diverging systems with $P_{\rm
f} \gg P_{\rm i}$, 
the converging systems with $P_{\rm f} \ll  P_{\rm i}$, and the
parallel systems with $P_{\rm f} \sim P_{\rm i}$. As an example, we
present the calculated results for a $1.4M_\sun+1.0M_\sun$ binary in
model 3, to illustrate the three kinds of evolutionary sequences in
Fig.~\ref{tp}. The corresponding bifurcation period is found to be
$1.25$ day, and the initial orbital periods are chosen to be $P_{\rm
i}=1.20$, $1.25$, and $1.40$ days, which represent the converging,
parallel, and diverging systems respectively.


\subsection{Effect of MB and mass loss}
\citet{pylyser88} have emphasized the effect of MB on the evolution
of LMXBs. Comparing the results of models 1 and 2 presented in
Table~\ref{tab1}, we find that the bifurcation periods with
traditional MB law are smaller (larger) than those with saturated MB
law, when the initial secondary star mass $M_{\rm 2,i}$ is less
(larger)
than $0.7M_\sun$. 


Our results suggest that mass loss also influences the value of
$P_{\rm bif}$, though in an less important way compared with MB. The
bifurcation periods in non-conservative models 3 and 4 are lower than
those in model 2, in which conservative mass transfer has been
assumed. This result is consistent with \citet{sluys05b}, but
contradicted with \citet{ergma98}.

It is also interesting to see whether an UCXB can form with
saturated MB. For an LMXB with an initial orbital period below the
bifurcation period, mass transfer is mainly driven by the loss of
angular momentum. The orbital period will decrease with the donor
mass until a minimum period is reached. \citet{paczynski81} found a
minimum period about $80$ min without MB, while \citet{pod02} showed
that minimum orbital periods less than $11$ min could be reached for
binaries with an initial orbital period very close to the bifurcation 
period if traditional MB is included, but in a time longer than the age of the 
universe. \citet{sluys05b} further investigated this ``magnetic capture" 
scenario for the formation of UCXBs. Our calculations show that when the 
initial orbital period is close to the bifurcation period, ultra-compact 
systems ($P<1$ h) can indeed form with saturated MB, but also in a time 
longer than the age of the universe. For example, for an LMXB with 
$M_{2,\rm i}=1.3M_\sun$ and $P_{\rm i}=0.46$ day in model 4, a final 
period of $P_{\rm f}=22$ min can be reached after $\ga 15$ Gyr of
mass transfer. All the works done by previous authors show that a more 
efficient angular momentum loss mechanism is required to produce UCXBs 
within $13.7$ Gyr in this scenario.

\subsection{Semi-analytical Method}
In this subsection we will try to use a semi-analytical method to
understand our numerical results. First from Eq.~(2), we have the 
following equation
\begin{equation}
\frac{3}{2}\frac{\dot{J}}{J}=\frac{3}{2}(\frac{\dot{M_1}}{M_1}+\frac{\dot{M_2}}{M_2})-
\frac{1}{2}\frac{\dot{M_1}+\dot{M_2}}{M_1+M_2}+\frac{1}{2}\frac{\dot{P}}{P}.
\end{equation}
If we assume a fraction $\alpha$ of the mass lost by the donor is
accreted by the NS, i.e., $\dot{M_1}=-\alpha\dot{M_2}$, we can write
the period derivative as
\begin{equation} \label{pdot}
\frac{\dot{P}}{P}=3\frac{\dot{J}}{J}-A(M_1,M_2,\alpha)\frac{\dot{M_2}}{M_2},
\end{equation}
where
\begin{equation}
A(M_1,M_2,\alpha)=\frac{3M_1^2+2(1-\alpha)M_1M_2-3\alpha
M_2^2}{M_1(M_1+M_2)}.
\end{equation}
Our analysis is limited to binary evolution with $M_2 < M_1$. In
this case it is clearly seen that mass transfer increases the
orbital period and angular momentum loss decreases the orbital
period. The bifurcation period is decided by the balance of these
two factors.

Keeping the orbital period unchanged (i.e. $\dot{P}\simeq 0$), we
calculate the maximum mass transfer rates for orbital periods from
$0.35$ day to $0.95$ day. This period interval covers the whole
range of the bifurcation periods obtained in this work and in
\citet{pod02}. We show the calculated mass transfer rates in
Fig.~\ref{mdot2}, and find that they can be fitted by an approximate
expression as
\begin{equation} \label{mdot}
-\dot{M_2}(P)\simeq 2.73\times 10^{-8} (P/{\rm day})^{6.41(\pm
0.11)}M_\sun \rm{yr}^{-1}.
\end{equation}
Then we calculate the mean mass transfer rates with constant $P$ and
find that they lie between $\dot{M}_2(P)/2$ and $\dot{M}_2(P)$. This
means that if we use Eq.~(\ref{mdot}) to calculate $P_{\rm rlof}$, it
will deviate no more than $\sim 10\%$ from the true value. The mean mass 
transfer rates here are calculated as follows. We fix the binary period $P$ in 
a constant value in our code, and then evolve the donor from its initial mass 
$M_{\rm 2,i}$ to the time when it loses half of its initial mass $0.5M_{2,i}$. 
This mass transfer process takes a time of $T_{\rm 1/2}$. Then we use 
$0.5M_{\rm2,i}/T_{1/2}$ as the mean $\dot{M}_2$ for this period $P$. 
If we assume that angular momentum loss is dominated by saturated MB when
$P<10$ d, from Eqs.~(2) and (\ref{newmb}) we have
\begin{eqnarray}
\frac{\dot{J}}{J_{\rm orb}} & = & -K \omega_{\rm crit}^2
(\frac{R_2}{R_{\sun}})^{0.5}(\frac{M_2}{M_{\sun}})^{-0.5}
(\frac{4\pi^2}{G})^{2/3}\frac{(M_1+M_2)^{1/3}}{M_1 M_2}P^{-4/3} \nonumber \\
&\simeq &  -6.2\times 10^{-11}(P/{\rm day})^{-1} {\rm yr}^{-1}.
\end{eqnarray}
Here we adopt  $K=2.7\times10^{47}$ gcm$^2$s \citep{andronov03},
$\omega_{\rm crit}=2.9\times 10^{-5}$ Hz \citep{sills00},
$M_1=1.4M_\sun$, $M_2=1M_\sun$, and replace the radius of the
secondary $R_2 $ with its Roche lobe radius $R_{\rm L,2}$ from
Eq.~(1). Combining Eqs.~(\ref{pdot})-(14) with $\dot{P}=0$ we obtain
$P_{\rm rlof}\simeq 12.4$ hr for conservative mass transfer
($\alpha=1$), and $P_{\rm rlof} \simeq 10.7$ hr for non-conservative
mass transfer ($\alpha=0$). These values agree well with our
numerical results ($\sim 10.8-11.1$ hr). If the traditional MB law
is used, similarly, from Eq.~(\ref{oldmb}) we get
\begin{eqnarray}
\frac{\dot{J}}{J_{\rm orb}} & = & -3.8\times 10^{-30}M_2R_2^4\omega^3
G^{-2/3}\frac{(M_1+M_2)^{1/3}}{M_1 M_2}\omega^{1/3} \nonumber \\
&\simeq &  -4.6\times 10^{-9}(P/{\rm day})^{-2/3} {\rm yr}^{-1},
\end{eqnarray}
for $M_1=1.4M_\sun$ and $M_2=1M_\sun$. Combining Eqs.~(\ref{pdot})-(13), (15) 
with $\dot{P}=0$ we get $P_{\rm rlof}\simeq22.3$ hr for conservative mass 
transfer ($\alpha=1$), which is about $20\%$ larger than $\sim 18.3$ hr 
from our numerical calculations and $\sim 17.7$ hr in \citet{pod02}. 
The main reason for this difference is that we use the constant value 
$1.4M_\sun$, $1M_\sun$ for $M_1$, $M_2$ in Eq.~(12), which should change 
with time to $\sim 2.2M_\sun$, $\sim 0.2M_\sun$. This will decrease 
the coefficient in Eq.~(12) and increase the value of $P_{\rm rlof}$ by 
$\sim 10-20\%$. For donor mass $\ge 1.2M_\sun$, this will increase 
the value of $P_{\rm rlof}$ by as much as $\sim 30\%$. So it is better to 
use $\alpha=0$ instead of $\alpha=1$ for donors mass $\ge 1.2M_\sun$, which could yield 
more accurate results (from our numerical results we find that the deviation 
of $P_{\rm rlof}$ between conservative and non-conservative mass transfer is 
smaller than $10\%$).

Using Eq.~(11)-(15), we also compute the semi-analytical results of
$P_{\rm rlof}$ for $0.5-2M_\sun$ donors, and compare them with our 
numerical results of models 1 and 4 in Fig.~4, where the semi 4 results 
are calculated with $\alpha=0$. When we calculate the semi 1 results in 
Fig.~4, for the reasons mentioned above and below, we use $\alpha=1$ for 
$M_{\rm 2,i} \le 1.1M_\sun$ and $\alpha=0$ for $M_{\rm 2,i} \ge 1.2M_\sun$, 
where $\alpha=1$ should be used. A few points need to be noted for the 
semi-analytical results in Fig.~4. Firstly with the above mentioned equations it 
is impossible to compute the $P_{\rm rlof}$ for binaries with $M_{\rm 2,i}\ge 1.4M_\sun$ 
under conservative mass transfer, since both terms in the right side of 
Eq.~(11) are negative when $\alpha=1$ and $M_{\rm 2,i}\ge1.4\,M_\sun$, 
and there will be no solutions for $\dot{P}=0$. We instead adopt $\alpha=0$ 
when $M_{\rm 2,i}>1.4M_\sun$ (from our numerical results we find that the 
deviation of $P_{\rm rlof}$ between conservative and non-conservative mass 
transfer is smaller than $10\%$). Secondly, Eq.~(13) is derived only for 
$1M_\sun$ donor star rather donors in the whole mass range ($0.5-2\,M_\sun$),
because in the latter case it is impossible to find a unified
expression of the mass transfer rate like Eq.~(13). As seen in
Fig.~4, the difference between the semi-analytical and numerical
results is generally smaller than $20\%$ except for donors smaller
than $0.7M_\sun$. The reasons for the big discrepancies when 
$M_{2,i}<0.7M_\sun$ are discussed in \S 4.

\section{Discussion and Conclusions}

Motivated by new ideas about MB and mass loss in LMXB evolution, we
have made a systematic investigation on the bifurcation periods in
binary models, taking into account different MB laws and mass loss
mechanisms. We find that the strength of MB is the dominant factor
in determining the value of bifurcation periods compared with mass
loss. The stronger MB, the larger the bifurcation periods. This also
results in an upper limit for the secondary masses beyond which no
converging systems exist.



In our calculations we assume either fully conservative (models 1
and 2) or non-conservative (models 3 and 4) mass transfer to
constrain the bifurcation period distribution in different mass
transfer modes. From the expression of $A(M_1,M_2,\alpha)$ we always
have
\begin{equation}
A(M_1,M_2,1)=3-\frac{3M_2}{M_1} <A(M_1,M_2,0)=
3-\frac{M_2}{M_1+M_2},
\end{equation}
which means that non-conservative mass transfer contributes more to
the increase of the orbital period than conservative mass transfer.
This explains why we generally have a lower bifurcation period in
non-conservative mass transfer models (models 3 and 4) than in
conservative mass transfer model (model 2) under the same MB law.
The real situation may lie between these two extreme cases. For
binary systems with donors $M_{\rm 2,i}\sim 0.5-0.8 M_\sun$, it
would take more that 13.7 Gyr before mass transfer begins via Roche
lobe overflow. So the bifurcation period for these system seems
meaningless, unless there exist some unknown mechanisms of loss of
orbital angular momentum.

\begin{figure}
\epsscale{.80}
\plotone{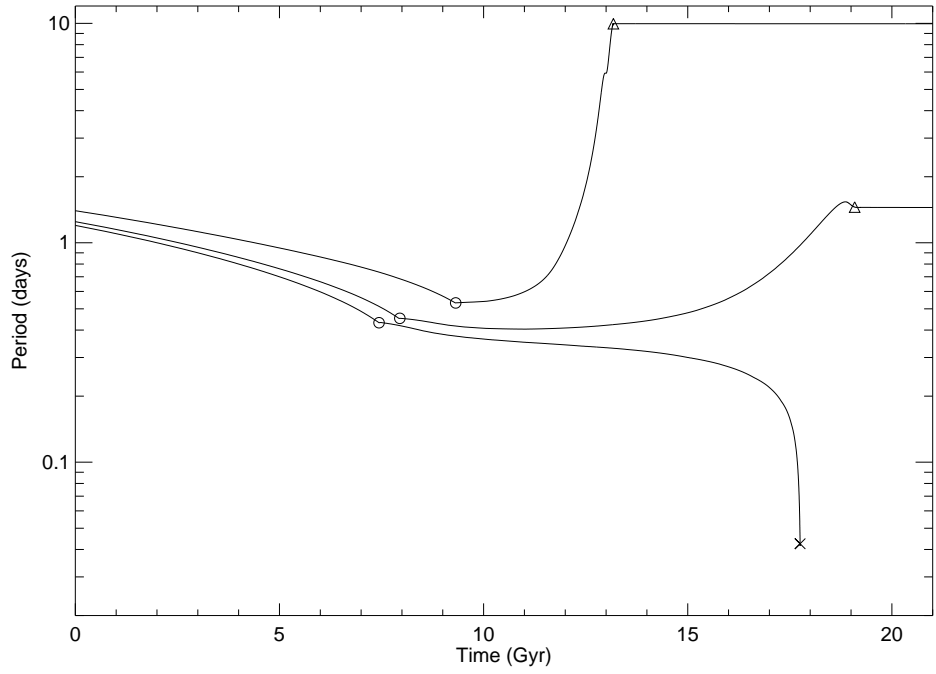}
\caption{Period evolutions of an LMXB with $M_{\rm 2,i}=1.0M_\sun$ and
$P_{\rm i}=1.20, 1.25, 1.40$ day in model 3. Circle, triangle, and cross
mark the onset of the Roche lobe overflow (RLOF), the end of the RLOF,
and the end of the calculation, respectively.
 \label{tp}}
\end{figure}

\begin{figure}
\epsscale{.80} \plotone{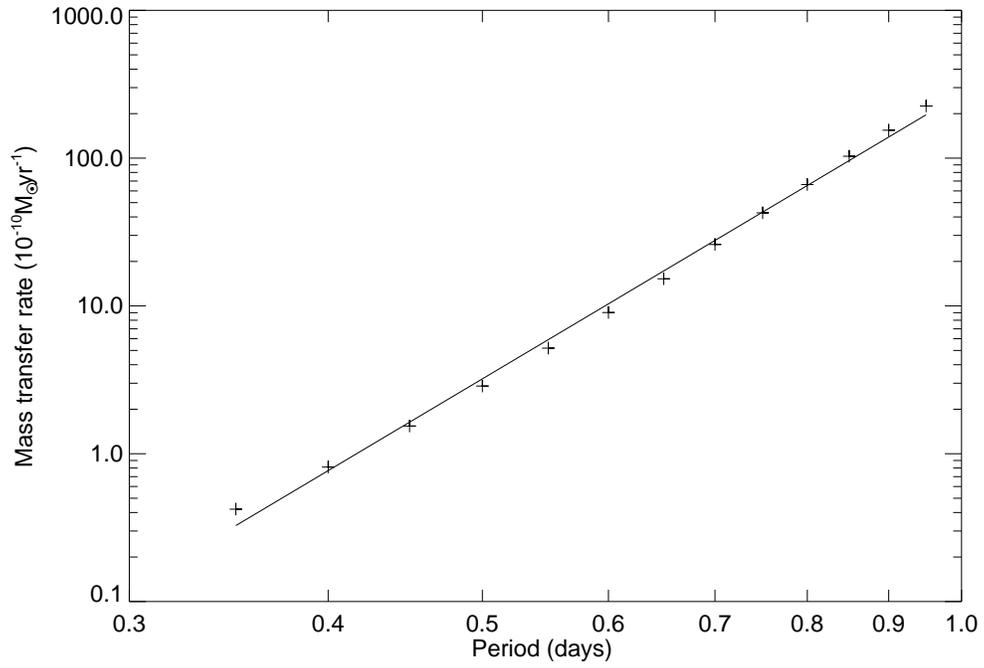} \caption{Maximum mass transfer
rates in an LMXB consisting of $1.4M_\sun$ neutron star and a
$1M_\sun$ secondary at fixed orbital periods from $0.35$ d to $0.95$
d. Crosses marks the calculated data and the solid line represents a
logarithmic fit. \label{mdot2}}
\end{figure}

In our semi-analytical analysis in \S 3.3 we use the condition
$\dot{P}\sim 0$ to derive the values of $P_{\rm rlof}$. This
expression seems different from  $P_{\rm f}\simeq P_{\rm rlof}$, which
is the original definition of bifurcation period. 
We argue here that these two expressions are roughly the same except for binaries 
with $M_{\rm 2,i}\ge1.4 M_\sun$ under conservative mass transfer ($\alpha=1$), 
the reason of which has been given in \S 3.3. For $M_{\rm 2,i}< 1.4 M_\sun$, 
we find that $\dot{P}/P$ always scales with $P$ from Eq.~(11) and (13)-(15). 
This means that if initially $\dot{P}>0$ ($<0$), $\dot{P}/P$ will become 
larger (smaller) during the evolution, leading to monotonic increase 
(decrease) of the period, as seen in Fig.~\ref{tp}. So for these systems  
$P_{\rm f} \sim P_{\rm rlof}$ is approximately equivalent with $\dot{P} \sim 0$. 
Several rough assumptions in this semi-analytical method contribute to the 
discrepancies between the semi-analytical results and the numerical results in Fig.~4, especially 
for $M_{\rm 2,i}<0.7M_\sun$. First is the use of $\dot{P}\sim0$ as the 
definition of $P_{\rm rlof}$, which may not work well sometimes. Second is the 
use of Eq.~(13), which is most suitable for binaries with $M_{2,i}=1M_\sun$ as 
pointed out in \S 3.3. Third is the assumption we made that magnetic braking 
law is the dominated  mechanism for the angular momentum loss, while the true 
case is that the MB may not work sometimes (for example when the convective envelop 
is too small). Fourth is that we use a constant initial value of $M_{\rm 1,i}$, 
$M_{\rm 2,i}$ for $M_1$, $M_2$ in Eq.~(12) and  Eq.(14)-(15), while in the true case 
$M_1$, $M_2$ should change with time. This will cause a big problem for $\alpha=1$ 
when $M_{\rm 2,i} > 1M_\sun$, which has been pointed out in \S 3.3. Fifth reason is 
the use of $\omega_{\rm crit,\sun}$ in Eq.~(14) as the value of $\omega_{\rm crit}$ 
for all the donors ranging from $0.5M_\sun$ to $1.3M_\sun$. At last we conclude 
that (1)$\dot{P}$ is a fair definition of bifurcation period, and (2)the period 
evolution during the mass transfer phase is in first approximation sufficiently 
well described by the balance of mass transfer and angular momentum loss caused 
by MB. For these rough assumptions made in this semi-analytical method, its 
results agree with the numerical results only in the leading order.


Our numerical calculations show that there is an upper limit for the
donor mass beyond which no converging systems will form.
\citet{pylyser88} found that, in the case of $M_{1, \rm i} =
4.0M_\sun$, there is no converging system existing if $M_{2,\rm i}>
1.7 M_\sun $, and concluded that for any given initial accretor mass
there exists a maximum initial secondary mass for the formation of
converging systems. From our calculations with $M_{1,\rm
i}=1.4M_\sun$, we find an upper limit for the initial secondary mass
$M_{2,\rm i}$ between $1.2$ and $1.4 M_\sun$ under saturated MB. 
The reason is that for binaries with a MS donor of initial mass 
$> 1.4M_\sun$, the bifurcation period is shorter than the minimum ZAMS 
period, so that these systems will diverge.
For traditional MB, this upper limit is $> 2M_\sun$, beyond the range of donor
masses we adopt.





\acknowledgments We are grateful to an anonymous referee for helpful
comments. This work was supported by the Natural Science Foundation
of China under grant numbers 10573010 and 10221001.  B.M. thank P.
P. Eggleton, Xuefei Chen, and Xiaojie Xu for their help during this
work.


\clearpage

\begin{deluxetable}{rrrrrrrrrr}
\tablecolumns{10} \tablewidth{0pc} \tablecaption{Calculated results
of the bifurcation periods for different binary models \label{tab1}}
\tablehead{ \colhead{}    &  \multicolumn{4}{c}{$P_{\rm bif}$ (day)}
& \colhead{}   &
\multicolumn{4}{c}{$P_{\rm rlof}$ (hr)} \\
\cline{2-5} \cline{7-10} \\
\colhead{$M_{2,\rm{i}}$ } & \colhead{Model 1}   & \colhead{ Model 2}    & \colhead{ Model 3} &
\colhead{ Model 4} &\colhead{}   & \colhead{ Model 1}   & \colhead{ Model 2}    & \colhead{ Model 3} &
\colhead{ Model 4}
}
\startdata
0.5$M_\sun$ &3.20 &4.08 &4.12 &4.01 & & 31.8 &16.5 &18.2 &13.6 \\
0.6$M_\sun$ &3.02 &3.41 &3.40 &3.35 & & 25.6 &11.3 &11.1 &10.0 \\
0.7$M_\sun$ &2.86 &2.81 &2.79 &2.77 & & 21.5 &10.0 &9.8 &9.6 \\
0.8$M_\sun$ &2.77 &2.41 &2.38 &2.37 & & 19.0 &10.1 &9.9 &9.8 \\
0.9$M_\sun$ &2.73 &2.14 &2.09 &2.10 & & 17.9 &10.7 &10.3 &10.4 \\
1.0$M_\sun$ &2.28 &1.29 &1.25 &1.27 & &18.3 &11.4 &10.8 &11.1 \\
1.1$M_\sun$ &1.51 &0.63 &0.59 &0.61 & &18.9 &11.7 &10.8 &11.2 \\
1.2$M_\sun$ &1.01 &0.55 &0.48 &0.51 & &19.1 &11.8 &10.5 &11.0 \\
1.3$M_\sun$ &0.86 &0.52 & &0.46 & & 19.2 &12.0 & &11.1 \\
1.4$M_\sun$ &0.90 & & & & &20.7 & & & \\
1.6$M_\sun$ &0.95 & & & & &22.3 & & & \\
1.8$M_\sun$ &0.98 & & & & &22.8 & & & \\
2.0$M_\sun$ &0.92 & & & & &21.6 & & & \\
\enddata
\end{deluxetable}

\clearpage

\clearpage

\begin{figure}
\epsscale{.80} \plotone{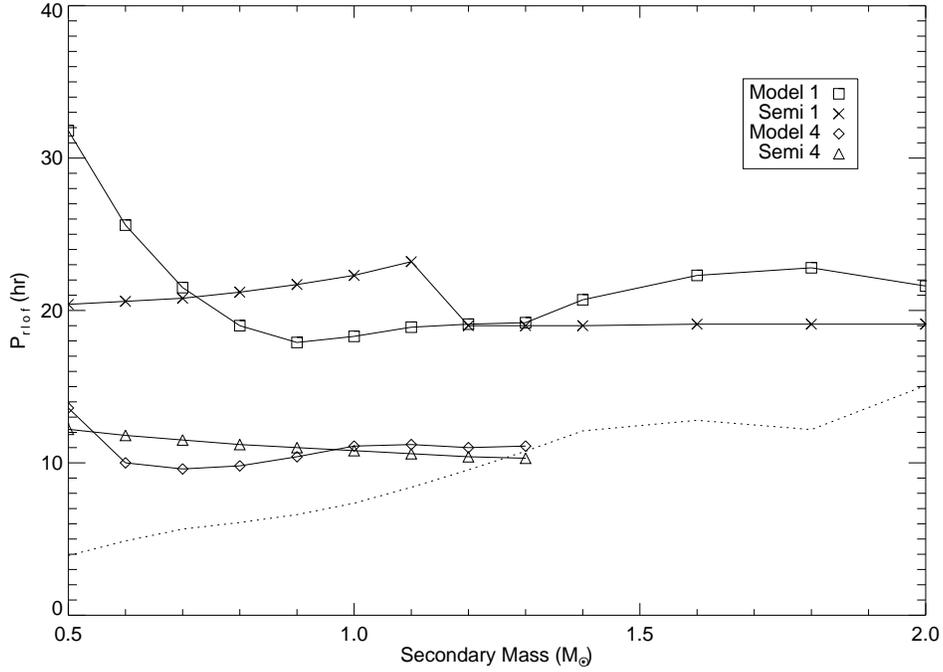} \caption{Comparison of the
semi-analytical results of $P_{\rm rlof}$ with the numerical values
for model 1 and model 4. The dotted line shows the minimum initial
period $P_{\rm ZAMS}$ that corresponds to a Roche lobe filling
zero-age main-sequence secondary star.  Here the semi 1 results 
are calculated with $\alpha=1$ for $M_{\rm 2,i} \le 1.1M_\sun$, and $\alpha=0$ 
for $M_{\rm 2,i} \ge 1.2M_\sun$ where $\alpha=1$ should be used. 
The reasons why we do this are given in the text. \label{bifurolf}}
\end{figure}

\clearpage

\clearpage

\end{document}